\documentclass[fleqn,10pt]{wlscirep}
\usepackage[utf8]{inputenc}
\usepackage[T1]{fontenc}
\usepackage{multicol}
\usepackage{subcaption}

\title{Gravitomagnetic Clock Effect: Using GALILEO to explore General Relativity}

\author[1, 4]{Jan Scheumann*}
\author[1, 4]{Dennis Philipp*}
\author[1, 4]{Sven Herrmann*}
\author[1, 4]{Eva Hackmann}
\author[1, 4]{Benny Rievers}
\author[2]{Javier Ventura-Traveset}
\author[3]{Luis Mendes}
\author[1, 4]{Claus Lämmerzahl}

\affil[1]{Center of Applied Space Technology and Microgravity (ZARM), University of Bremen, 28359 Germany}
\affil[2]{Centre Spatial de Toulouse, European Space Agency, 18 Avenue Edouard Belin, CEDEX 9, 31401 Toulouse, France}
\affil[3]{Rhea Group for ESA, ESAC, Villanueva de la Cañada, Madrid, 28692 Spain}
\affil[4]{Gauss-Olbers Space Technology Transfer Center, c/o ZARM, University of Bremen, 28359 Germany}

\affil[*]{jan.scheumann@zarm.uni-bremen.de}
\affil[*]{dennis.philipp@zarm.uni-bremen.de}
\affil[*]{sven.herrmann@zarm.uni-bremen.de}

\begin{abstract}
All experiments to date are in remarkable agreement with the predictions of Einstein's theory of gravity, General Relativity.
Besides the classical tests, involving light deflection, orbit precession, signal delay, and the gravitational redshift, modern technology has pushed the limits even further. 
Gravitational waves have been observed multiple times as have been black holes, arguably amongst the most fascinating objects populating our universe. 
Moreover, geodetic satellite missions have enabled the verification of yet another prediction: gravitomagnetism.
This phenomenon arises due to the rotation of a central body, e.g., the Earth, which is dragging spacetime along. 
One resulting effect on satellite orbits is the observed Lense-Thirring effect.
Another predicted, yet unverified, effect is the so-called gravitomagnetic clock effect, which was first described by Cohen and Mashhoon as the proper time difference of two counter-revolving clocks in an orbit around a rotating mass. 

A theoretical framework is introduced that describes a gravitomagnetic clock effect based on a stationary spacetime model. An incremental definition of a suitable observable follows, which can be accessed via orbit data obtained from the European satellite navigation system \textit{Galileo}, and an implementation of the framework for use with real satellite and clock data is presented. 
The technical requirements on a satellite mission are studied to measure the gravitomagnetic clock effect at the state-of-the-art in satellite laser ranging and modelling of gravitational and non-gravitational perturbations. 
Based on the analysis within this work, a measurement of the gravitomagnetic clock effect is highly demanding, but might just be within reach in the very near future based on current and upcoming technology.
\end{abstract}
\begin{document}

\flushbottom
\maketitle
\thispagestyle{empty}
\section*{Introduction}

All experiments to date are in remarkable agreement with the predictions of Einstein's theory of gravity, General Relativity (GR). However, there are several reasons to assume that GR  needs to be modified, among them the inevitable appearance of singularities, the inability to renormalize GR, and the rising tension in cosmology.
Therefore, it is of utmost importance for fundamental physics to experimentally test the predictions and the underlying principles of GR. The classical tests involve the deflection of light, precession of the periastron, signal delay, and the gravitational redshift. These have been tested on Earth, in the Solar System, and also in astrophysical observations. Gravitational waves as predicted by GR have been observed indirectly and directly.\cite{will1919MeasurementDeflection2015, bertotti2003test,everittGravityProbeTest2015,vessotTestRelativisticGravitation1980}
From the large range of tests of GR two particular aspects are to be emphasized here: 

Firstly, modern technology enables to operate highly stable and accurate clocks in space environments. 
Via the comparison of at least two (standard) clocks it is possible to test the gravitational redshift, which is particularly important for precise navigation and positioning, e.g., as part of Global Navigation Satellite Systems (GNSS). 
Recently two \textsl{Galileo} satellites, which are on  eccentric orbits due to an injection failure, have been used to improve the thirty years old bounds on the test of the gravitational redshift.\cite{herrmannTestGravitationalRedshift2018,delvaNewTestGravitational2019} 
This improvement was mainly possible due to the on-board high performance atomic clocks based on passive hydrogen masers (PHM)\cite{berthoudEngineeringModelSpace2003}.

Secondly, in GR the source of the gravitational field is not only the mass density but also every other kind of energy as well as fluxes thereof. 
For a rotating body such as the Earth this leads to the so-called frame dragging. 
Another terminology, in analogy to electrodynamics, characterizes spacetime effects due to mass/energy currents as \textit{gravitomagnetic}. 
Consequences include the Lense-Thirring effect \cite{LenseThirring18,mashhoonGravitationalEffectsRotating1984}, i.e., a precession of the orbital plane (in the weak gravitational field), that has been measured by the Lageos/Lares satellite missions \cite{Ciufolini_1998,ciufoliniFundamentalPhysicsGeneral2013, ciufolini_test_2016}. 
Another consequence is the Schiff effect \cite{Schiff1960,Schiff1960b}, i.e., the spin-spin coupling, resulting in a precession of a gyroscope's spin in orbit around the Earth, as measured by the Gravity Probe B satellite mission \cite{Everittetal2011,everittGravityProbeTest2015}. 

However, frame dragging does not only act on orbital motion or spins, but also on clocks. 
The physical origin is the inﬂuence of the source’s spin dipole moment on the length of four-dimensional (4d) worldlines, i.e., trajectories through spacetime. 
This 4d length is called proper time, and a clock that shows proper time is called a standard clock \cite{perlickCharacterizationStandardClocks1987}. 
The so-called gravitomagnetic clock effect was first described by Cohen and Mashhoon as the proper time difference of two counter-revolving clocks on an idealized equatorial circular orbit around a rotating mass \cite{cohen_standard_1993}. 
It is a genuine general relativistic effect, and remains yet unverified. 
The natural question arises if state-of-the-art or near-future technology might be sufficient to close this gap, and if GR will once more hold up to all tests. 
In this paper we provide a theoretical framework to analyse this question and provide a first estimate to its answer.

Following the first definition of Cohen and Mashhoon\cite{cohen_standard_1993} a number of related but slightly different definitions of gravitomagnetic clock effects have been developed in the literature \cite{Binietal2001, mashhoon_measuring_1998, Tartaglia2000, Mashhoonetal2001, hackmann_generalized_2014, iorioReviewGravitomagneticClock2023}. 
One may unify these approaches as follows:

\smallskip
\emph{A gravitomagnetic clock effect (GMCE) is a (linear) combination of the elapsed proper times on two distinct worldlines, corrected for all gravitoelectric contributions such that the remaining observable depends, to first order, on the angular momentum of the gravitating source.}
\smallskip

Usually, in the literature these worldlines are considered to be (geodesic) satellite orbits. 
Two events on each worldline are chosen. 
Following Cohen and Mashhoon \cite{cohen_standard_1993}, these may correspond to successive transits of an azimuthal position, e.g., when each satellite closes the azimuthal angle from $\varphi = 0$ to $\varphi = \pm 2\pi$. 
Other choices include events with fixed coordinate times \cite{mashhoon_measuring_1998} or at satellite meeting points \cite{Tartaglia2000}.
In this paper, we choose fixed (incremental) azimuthal positions. Then, the difference of elapsed proper times is labelled the (observer-dependent, two-clock) gravitomagnetic clock effect \cite{Binietal2001}. 
Note that of course the elapsed proper times are invariant, but the notion of the azimuthal arcs depend on the chosen observer \cite{BonnorSteadman1999}.

The GMCE studied by Cohen and Mashhoon, i.e., the proper time difference of two counter-revolving clocks on circular orbits of the same radius in the equatorial plane of the Earth, is of the order of $10^{-7}$ seconds per revolution, independent of the radius of the two orbits. 
A generalized definition in the framework of the parametrized post-Newtonian formalism, including, e.g., the effects of the nonspherical shape of the Earth, was considered\cite{Mashhoonetal2001}. 
Eccentric and inclined orbits were discussed\cite{mashhoon_measuring_1998, Mashhoonetal2001}, as well as arbitrary geodesic orbits\cite{hackmann_generalized_2014}. 
A generalisation for arbitrary worldlines/satellite orbits, however, is still missing. 
A dedicated satellite mission to measure the GMCE effect of Cohen and Mashhoon was proposed by Gronwald and others\cite{gronwald_gravity_1997}. 
Gravitational and non-gravitational error sources for such a mission have been discussed \cite{gronwald_gravity_1997,Iorio2001,Iorio2001b,Iorioetal2005,Lichteneggeretal2006}. 
From these studies, it can be concluded that the most challenging task for a mission to measure a GMCE is not the stability and accuracy of the orbiting clocks but the precise tracking of the satellites. 
This is needed because of the imperfect cancellation of the Keplerian periods of the two clocks, which induces large errors in the measurements.

In this paper we investigate i) a theoretical model to construct a suitable definition of a GMCE and the related observables and ii) a source for reliable data that could be feasible to experimentally detect the GMCE. 
Concerning (i), we construct a general, incremental definition for arbitrary worldlines and compare it to the approaches in the literature. 
For (ii), a promising source for experimental data is given by the ensembles of clocks in GNSS constellations, as they have already proven to be useful in other tests of GR \cite{herrmannTestGravitationalRedshift2018, delvaNewTestGravitational2019}.

The theoretical framework is presented in the next section, starting from the initial descriptions of ideal orbits up to a definition suitable for numerical data analysis. 
Thereupon, the potential use of \textsl{Galileo} satellite data for a measurement of the GMCE is outlined and technical requirements and challenges for a dedicated mission are analysed.

\section*{Theoretical Model}

\subsection*{First theoretical descriptions}

Mashhoon and Cohen\cite{cohen_standard_1993} observed that the elapsed proper time of a clock after a full $2\pi$ pro- (+) or retrograde (-) revolution on a circular equatorial orbit in the Kerr spacetime is
\begin{equation}
    \tau_{\pm} = \frac{2\pi}{\omega_0}\sqrt{1- \frac{3 G M}{c^2 r} \pm 2\frac{J}{M c^2}\omega_0} \approx \frac{2\pi}{\omega_0} \left( 1- \frac{3 G M}{2 c^2r} \pm \frac{J}{M c^2} \omega_0 \right),
\end{equation}
with $\omega_0 = \sqrt{\frac{GM}{r^3}}$ denoting the Kepler period of the orbit. 
Here, $M$ is the mass of the Earth and $J$ its angular momentum.
The difference between the pro- and retro-grade result is
\begin{equation}
\tau_+ - \tau_- \approx  \frac{4 \pi J }{M c^2} \, ,
\end{equation}
which is independent of $G$ and $r$ at leading order but directly proportional to the angular momentum of the central object.. 
This difference has been termed the gravitomagnetic clock effect since then. 

To allow for a measurement of gravitomagnetism with clocks, the result has to be generalized beyond the case of identical circular orbits.
This has been done for the Kerr spacetime and arbitrary geodesic orbits of inclination $i$, eccentricity $e$ and semi-major axis $ a$.\cite{hackmann_generalized_2014}
In this case, at leading order the proper time for a full $2\pi$ revolution is
\begin{equation}
    \tau_{\pm} = \frac{2\pi}{\omega_0} \left( 1- \frac{(1+e^2)}{(1-e^2)}\frac{3G M}{2c^2 a} \right) \pm \frac{2\pi J}{M c^2} \frac{\cos i (3e^2+2e+3)-2e-2}{(1-e^2)^{\frac{3}{2}}},
    \label{eq:generalized_tau}
\end{equation}
where $\omega_0 = \sqrt{\frac{GM}{a^3}}$ is the Keplerian orbital period again. 
The proper time difference for pro- and retro-grade rotation on identically shaped orbits leads to
\begin{equation}
    \tau_+ - \tau_-  \approx \frac{4\pi J}{M c^2}\frac{\cos i (3e^2+2e+3)-2e-2}{(1-e^2)^{\frac{3}{2}}} \, ,
\end{equation}
which reduces to the result above for a circular equatorial orbit. 
The difference in proper times between two not necessarily identical orbits is
\begin{equation}
\Delta \tau_{\text{gm}} \approx \frac{2\pi J}{Mc^2} \left[
    s_1 \frac{\cos i_1 (3e_1^2+2e_1+3)-2e_1-2}{(1-e_1^2)^{\frac{3}{2}}}
    - s_2 \frac{\omega_{02}}{\omega_{01}} \frac{\cos i_2 (3e_2^2+2e_2+3)-2e_2-2}{(1-e_2^2)^{\frac{3}{2}}}       \right],
\label{eq:d_tau_first_order}                                  
\end{equation}
where, $e_{\text{i}}, i_{\text{i}}, \omega_{0\text{i}}$ are defined for the orbits $i = 1,2$ as above for equation (\ref{eq:generalized_tau}) and $s_{\text{j}}$ equals to $\pm 1$ depending on the revolution sense. 
Note that all gravitoelectric contributions cancel out, i.e., the time differences vanishes for $J=0$.

Obviously, a measurement of this effect relies on the exact knowledge of the orbital parameters. 
At the same time these orbits are treated as ideal geodesic orbits in a Kerr spacetime and mass multipole moments of the Earth as well as non-gravitational disturbances are not yet included. 
However, any satellite is subject to these effects. 
Furthermore, orbital data is mostly given in a sampled, incremental form and not via ideal Keplerian orbital elements. 

\subsection*{Incremental definition of an observable}
To obtain a more suitable definition of the GMCE that can be applied also to incremental data and orbits that deviate significantly from ideal Kepler ellipses, we reformulate a GMCE observable and use a first-order post-Newtonian framework.

\subsubsection*{ppN spacetime}

A parametrized post-Newtonian (pN) spacetime metric of the form
\begin{align}
    g = g_{tt} dt^2 + 2g_{ti} dt dx^i + g_{ij} dx^i dx^j \, ,
\end{align}
is used to describe the Earth's exterior, in which  
\begin{subequations}
\begin{align}
    g_{tt} &= - \left(1 + 2U/c^2 + 2\beta U^2/c^4 \right) c^2 \, ,\\
    g_{ti} &= -(\gamma+1) G\, \dfrac{(\vec{J} \times \vec{x})_i}{c^2r^3} \, \\
    g_{ij} &= \delta_{ij} \left(1-2 \gamma U/c^2 \right) \, .
\end{align}
\end{subequations}
Note that the coordinates $(t,x^i)$ are harmonic and $t$ has the SI-unit of a time.
We deliberately include only the ppN parameters $\beta$ and $\gamma$  and $\vec{J}$ is the Earth's angular momentum. 
For GR, the parameters values are $\beta =1$ and $\gamma=1$.
Note that for a rigid model of the Earth, which we shall assume for the following, the other pN potentials contributing to $g_{tt}$ at the $c^{-4}$-level can be included into the $U^2$-term and also the other vector potential contributing to $g_{ti}$ is absorbed.\cite{weinbergGravitationCosmologyPrinciples1972, soffelRelativityAstrometryCelestial1989}

This effectively means to include non-relativistic contributions to the energy density (internal, thermal, gravitational) into the mass density as well \footnote{via $1/c^2$ the energy densities are transformed into a mass density}.
Then, $U$ is a gravitational potential, which can be decomposed according to
\begin{align}
    U = - \dfrac{GM}{r} \left( 1 +  \sum_{l=2}\sum_{m=-l}^l (C_{lm} \cos(m\varphi) + S_{lm} \sin (m\varphi)) \left( \dfrac{R_0}{r} \right)^l P_{lm}(\cos \vartheta) \right) \, ,
\end{align}
where $P_{lm}$ are the associated Legendre functions of degree $l$ and order $m$, $C_{lm}$ and $S_{lm}$ are the mass multipole moments, $R_0$ is a reference radius, e.g., Earth's mean radius, and associated spherical coordinates $r,\vartheta,\varphi$ are employed. Note that, strictly, the moments are not the Newtonian ones.
In the limit $r\to \infty$ the potential $U$ vanishes. 
Thus, the coordinate time $t$ is the proper time of inertial observers infinitely far away.

Frame dragging effects due to the Earth's rotation are included via the angular momentum. 
For, e.g., a rigid sphere that may serve as a lowest-order approximation we have
\begin{align}
    J = I\, \Omega = \dfrac{2}{5}\, R_0^2\, M \, \Omega \, ,
\end{align}
where $R_0$ is the mean radius of the Earth and $ \Omega $ its angular velocity.

The formal pN-coordinate time $t$ is inaccessible for practical measurements. 
For real-world observations, we need to refine the spacetime model.
To this end, we introduce the proper time of a clock on the Earth's geoid $T$ \cite{philippDefinitionRelativisticGeoid2017, philippRelativisticGeoidGravity2020} as the new coordinate time. 
It is defined as the \textit{International Atomic Time}~(TAI) nowadays. 
For the proper time of clocks rigidly rotating on the geoid we have
\begin{align}
    -c^2 dT^2 = -c^2 e^{2\phi_0} dt^2 \quad \Rightarrow dT = e^{\phi_0} dt =: \left( 1 + \dfrac{U^*_0}{c^2} \right) dt \,,
\end{align}
with $ \phi $ being a time-independent redshift potential,
\begin{align}
    e^{2\phi} = -\frac{1}{c^2} \left( g_{tt} + 2 g_{t\varphi} \Omega + g_{\varphi\varphi} \Omega^2 \right) \, ,
\end{align}
that defines a relativistic gravity potential $U^*$.\cite{philippDefinitionRelativisticGeoid2017, philippRelativisticGeoidGravity2020, philippChronometricHeightRelativistic2021} 
On the geoid, the potential value is $\phi_0$, leading to a value of $ U^*_0 $, which must be agreed upon by convention. 
In the weak field limit $U^*$ of course reduces to the Newtonian gravity potential usually denoted by $W$. 
Therefore, a suitable choice is
\begin{align}
    U^*_0 = W_0 \, ,
\end{align}
where $W_0$ is the value characterizing the Newtonian geoid.\cite{sanchezConventionalValueGeoid2016} 
At the first pN order, we then have the components of the metric in using spherical coordinates:
\begin{subequations}
    \begin{align}
        g_{TT} &= - c^2 \left(1 + 2U/c^2 + 2\beta (U/c^2)^2 -2 U^*_0/c^2 + 3 (U^*_0/c^2)^2 - 4U U^*_0 /c^4 \right) \\
        g_{T\varphi} &= -(\gamma+1) \dfrac{G}{c^2} \dfrac{J}{r} \sin^2\vartheta \\
        g_{rr} &= \left(1-2\gamma U/c^2 \right) \, \\
        g_{\vartheta\vartheta} &= r^2 \left(1-2\gamma U/c^2 \right) \, \\
        g_{\varphi\varphi} &= r^2 \sin^2\vartheta \left(1-2\gamma U/c^2 \right) \,
    \end{align}
\end{subequations}
The components can also be formulated using Cartesian coordinates, which might be more suitable for some applications. The $g_{TT}$ term remains unchanged, while the other components read
\begin{subequations}
    \begin{align}
        g_{Ti} &= -(\gamma+1) G \dfrac{(\vec{J} \times \vec{r})_i}{c^2 \cdot r^3}\\
        g_{ij} &= \delta_{ij} \left(1-2\gamma U/c^2 \right)        
    \end{align}
\end{subequations}
The analysis of the equation of motion\footnote{It follows also directly from the so-called clock postulate.}, cf. Suppl.\ Equations (\ref{supp:eq:geodesic}) to (\ref{supp:eq:integral}), yields an integral formulation of a satellite's proper time for parametrization by the azimuthal angle $\varphi$ (assuming $\tau = 0$ at $\varphi = \varphi_0$) or by using Cartesian coordinates and a parametrization by TAI (assuming $\tau = 0$ at $T = T_0$),
\begin{subequations}
    \begin{align}
        c \tau(\varphi) &= \pm \int_{\varphi_0}^{\varphi} \sqrt{-g_{TT} \left( \dfrac{dT}{d\varphi} \right)^2 - 2 g_{T\varphi} \dfrac{dT}{d\varphi} - g_{ij} \dfrac{dx^i}{d\varphi} \dfrac{dx^j}{d\varphi} } \, d\varphi \, ,
        \label{integral_phi} \\
        c \tau(T) &= \pm \int_{T_0}^T \sqrt{-g_{TT} - 2 \left(g_{Tx} \dfrac{dx}{dT} + g_{Ty} \dfrac{dy}{dT} \right) - g_{ij} \dfrac{dx^i}{dT} \dfrac{dx^j}{dT} } \, dT \, . \label{integral_TAI}
    \end{align}
\end{subequations}
The sign needs to be chosen according to the direction of revolution.
Up to here, the results are general and assume no special type of orbit.

\subsubsection*{Defining the observable}

Only in the ideal cases mentioned directly measured quantities can be defined by the proper time difference after full azimuthal closure on either orbit. 
Taking this unique combination, all gravitoelectric effects cancel and the result is to be attributed to gravitomagnetism alone.
We infer that the GMCE in this case is a topological effect and $\Delta \tau$, as defined in equation~(\ref{eq:d_tau_first_order}), is a direct observable.
In the general situation, however, the difference in two proper times is not purely attributed to gravitomagnetic effects, but contains Newtonian and gravitoelectric parts.
Thus, a direct GMCE observable can not be defined and we have to pay attention to indirect, i.e., derived  observables.
For sure, any definition of ``the'' GMCE should ensure that all gravitoelectric contributions are canceled and for $J \to 0$ this GMCE vanishes. 
Moreover, for a good definition we require that the topological effect is recovered in the limit of circular geodesics.

One suggested definition\cite{hackmann_generalized_2014} is
\begin{align}
    ^{\text{HL}}{\Delta \tau}{_{\text{gm}}} (\pm 2 \pi) := \tau_1(\pm 2\pi) + \alpha \tau_2(\pm2\pi) \, , \quad \alpha = - \dfrac{\hat{\tau}_1(\pm2\pi)}{\hat{\tau}_2(\pm2\pi)} \, , \quad \hat{\tau}_i := \tau_i \big|_{J=0} \, .
\end{align}
Here, $\alpha$ ensures that gravitoelectric parts cancel and the effect vanishes for $J=0$. 
It is determined by the orbital elements of both orbits, see Suppl.~Equation~(\ref{supp:eq:d_tau_first_order}).
However, for real satellite orbits, the orbital elements are not constant, but Keplerian approximations hold only in the neighborhood of the momentary position.
Another possible definition of a GMCE is 
\begin{align}
    ^{\text{2}}{\Delta \tau}{_{\text{gm}}} (\pm 2 \pi) &:= \big( \tau_1 (\pm 2\pi) - \hat{\tau}_1(\pm 2\pi) \big) - \big( \tau_2 (\pm 2\pi) - \hat{\tau}_2(\pm 2\pi) \big)  \notag \\
    &= \big( \tau_1 (\pm 2\pi) -\tau_2 (\pm 2\pi) \big) - \big( \hat{\tau}_1 (\pm 2\pi) - \hat{\tau}_2 (\pm 2\pi)  \big) \, .
\label{eq:def2}
\end{align}
Note that this combination is anti-symmetric under the change of clock labels.

Instead of using the full azimuthal closure, this second definition can also be done in terms of arbitrary azimuthal angle intervals $\varphi_1$ and $\varphi_2$ related to the elapsed angle on the two orbits, respectively,
\begin{align}
    ^{\text{2}}{\Delta \tau}{_{\text{gm}}} (\varphi_1, \varphi_2) = \big( \tau_1 (\varphi_1) -\tau_2 (\varphi_2) \big) - \big( \hat{\tau}_1 (\varphi_1) - \hat{\tau}_2 (\varphi_2)  \big) \, . \label{eq:definition2}
\end{align}
Yet another dimensionless definition is
\begin{align}
    ^{\text{3}}{\Delta \tau}{_{\text{gm}}} (\varphi_1, \varphi_2) := \dfrac{\tau_1(\varphi_1)}{\hat{\tau}_1(\varphi_1)} - \dfrac{\tau_2(\varphi_2)}{\hat{\tau}_2(\varphi_2)} \, ,
\end{align}
which is again anti-symmetric.
All these definitions introduce an indirect observable which requires precise orbit data since only the true proper times $\tau_{i}$ are measured, but the $\hat{\tau}_i$ need to be modeled. 
Let us evaluate all three definitions for the special case of two circular geodesic orbits (with different radii) after full azimuthal closure on each orbit.
We have, for propagation directions indicated by $s_i = \pm 1$,
\begin{align}
    ^{\text{HL}}{\Delta \tau}{_{\text{gm}}} (\pm 2\pi) &= \dfrac{2\pi J }{Mc^2} \left( s_1 - s_2 \left( \dfrac{r_1}{r_2} \right)^{3/2} \right) \, , \\
    ^{\text{2}}{\Delta \tau}{_{\text{gm}}} (\pm 2\pi,\pm 2\pi) &= \dfrac{2\pi J}{Mc^2} (s_1 - s_2) \, , \\
    ^{\text{3}}{\Delta \tau}{_{\text{gm}}} (\pm 2\pi,\pm 2\pi) &= \dfrac{J}{M c^2} \left( s_1 \sqrt{\dfrac{GM}{r_1^3}} - s_2 \sqrt{\dfrac{GM}{r_2^3}} \right) \, .
\end{align}
If the radii coincide, all definitions reduce to the same topological effect described above and the numerical results of the first two definitions match. The third definition introduces a dimensionless result by dividing by the Kepler period. However, for different radii the three definitions give numerically different GMCE observables, with the first definition diverging for large $r_1$.
Though there is no naturally given choice of which observable to use, it seems that the second definition has some advantages:
It reproduces the topological effect, has the dimension of a time, does not diverge if one of the radii becomes large, and is antisymmetric under the exchange of clock labels.

To illustrate the different numerical values that follow from the definitions, we calculate the result for one geostationary orbit at $r\approx 36000\,$km and one MEO at $r\approx 1000\,$km revolving in pro- and retrograde directions, respectively.
We find
\begin{subequations}
\begin{align}
    ^{\text{HL}}{\Delta \tau}{_{\text{gm}}} (\pm 2\pi) &= 1.8 \cdot 10^{-5} \, \text{s} \\
    ^{\text{2}}{\Delta \tau}{_{\text{gm}}} (2\pi,-2\pi) &= 1.6 \cdot 10^{-7} \, \text{s} \, .
\end{align}
\end{subequations}
It is apparent, that different definitions of the observable can produce results that differ by several orders of magnitude, while still describing the same effect. Care must be taken when comparing different results as to which observable was used.

For the second choice of a GMCE, we define
\begin{align}
   \dfrac{ ^{\text{2}}{d\tau}{_{ \text{gm} } } }{d\varphi} &:= \left( \dfrac{d\tau_1}{d\varphi} - \dfrac{d\tau_2}{d\varphi} \right) - \left( \dfrac{d\hat{\tau}_1}{d\varphi} - \dfrac{d\hat{\tau}_2}{d\varphi} \right) \, ,
\end{align}
and recover the integrated effect after the same azimuthal increment,
\begin{align}
    ^{\text{2}}{\tau}{_{ \text{gm} } } (\varphi) &= \int_0^{T_{\text{gm}}} {}^{\text{2}}{d\tau}{_{ \text{gm} } } 
    = \int_0^\varphi \left( \left( \dfrac{d\tau_1}{d\varphi} - \dfrac{d\tau_2}{d\varphi} \right) - \left( \dfrac{d\hat{\tau}_1}{d\varphi} - \dfrac{d\hat{\tau}_2}{d\varphi} \right) \right) d\varphi 
    = \big( \tau_1(\varphi) - \hat{\tau}_1(\varphi) \big) - \big( \tau_2(\varphi) - \hat{\tau}_2(\varphi) \big) \, .
    \label{eq:gmce-def}
\end{align}
We choose this definition in the following.

\section*{Estimating the GMCE from GNSS data and orbit simulations}

For the purpose of this work we make use of clock and orbit products processed and supplied by the ESOC Navigation Support Office. 
This data spans from January 1st 2021 to March 31st 2021. 
During this time an independent ILRS measurement campaign on dark matter search took place, resulting in an increased number of laser ranging measurements on \textsl{Galileo} satellites\cite{bertrandGAlileoSurveyTransient2021}, thus improving the precision of the orbit products. 
For tables and graphs that show results for a single orbit or a single day from this data, we choose GPS week 2139, day 0 (January 3rd, 2021) as a representative example. 
The satellites that we look at in particular are E24 and E26, nominal \textsl{Galileo} satellites using a PHM as their respective master clock during this period. 

\subsubsection*{Clock proper time from orbit products}

\begin{figure}
    \centering
        \includegraphics[width=0.6\textwidth]{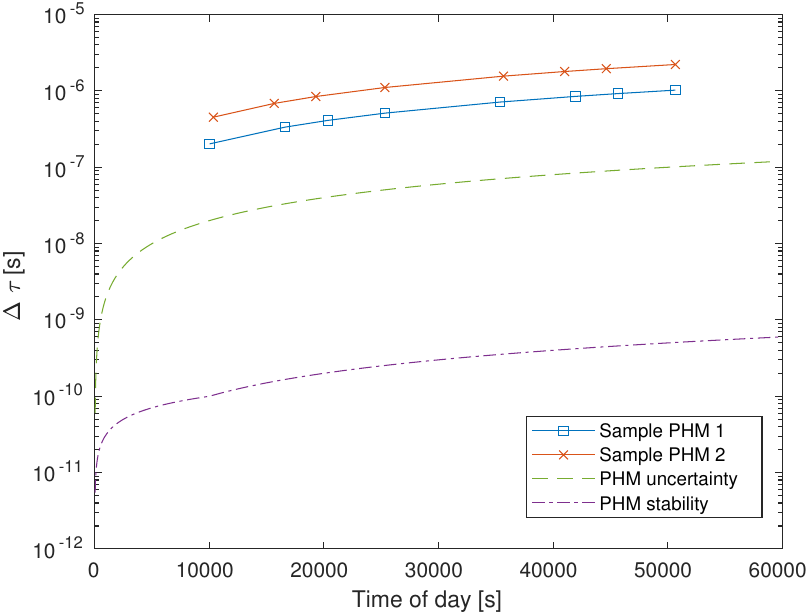}
       \caption{The two top lines show the difference between calculated and observed elapsed proper time on the satellite clock for two sample \textsl{Galileo} satellites, over one full orbit. The middle, dashed line represents the uncertainty in the absolute PHM clock frequency offset of $\Delta \nu/\nu < 2 \cdot 10^{-12}$. The lower line depicts the uncertainty due to clock frequency stability and is derived from the Allan variance as given in the \textsl{Galileo} PHM specifications.}
	\label{fig:deltatauAllanVar}
\end{figure}

To evaluate equation (\ref{integral_phi}) for some orbital arc, start and end points are picked at certain time stamps of the clock and orbit products, separated by $\Delta \varphi$. 
For the generalized definition of the GMCE any value of $\Delta \varphi$ can be used. 
We consider intervals of approximately $\Delta \varphi = \pi/8, \pi/4, \pi/2, \pi$ and $2\pi$ and compare the results to the time interval as determined from the clock products at the corresponding time stamps. 
As the on-board clocks of the \textsl{Galileo} satellites are detuned before launch to compensate relativistic gravitoelectric effects on the nominal circular orbit, we omit those contributions in the evaluation. 
Thus, the observed difference then reduces to 1\,$\mu$s per orbit and below as shown in Fig.\ \ref{fig:deltatauAllanVar}. 

This difference is still about one to two orders of magnitude above the expected magnitude of the gravitomagnetic contributions we are looking for. To leading order, this difference increases linearly in time. 
The most probable cause for this lies in (i) a deliberate (yet unknown to us) deviation of the frequency offset that is applied to the PHM pre-flight to compensate relativistic effects on a circular orbit and (ii) the uncertainty of the absolute PHM clock frequency $\nu$, which is about $\Delta\nu/\nu \approx 2 \times 10^{-12}$, see Fig.\ \ref{fig:deltatauAllanVar}.
This is the level of reproducibility of the PHM frequency when it is switched off and on again during launch and deployment. 
With respect to the observed clock drift during one orbit, (ii) translates into an uncertainty of about 100\,ns per orbit. 
This is a factor of 10 below the actually observed difference, but could be reached once the exact deliberate frequency offset for (i) is known.

\subsubsection*{Modelling the gravitomagnetic contribution}

\begin{figure}
    \centering
    \begin{subfigure}{0.4\textwidth}
        \includegraphics[width=\textwidth]{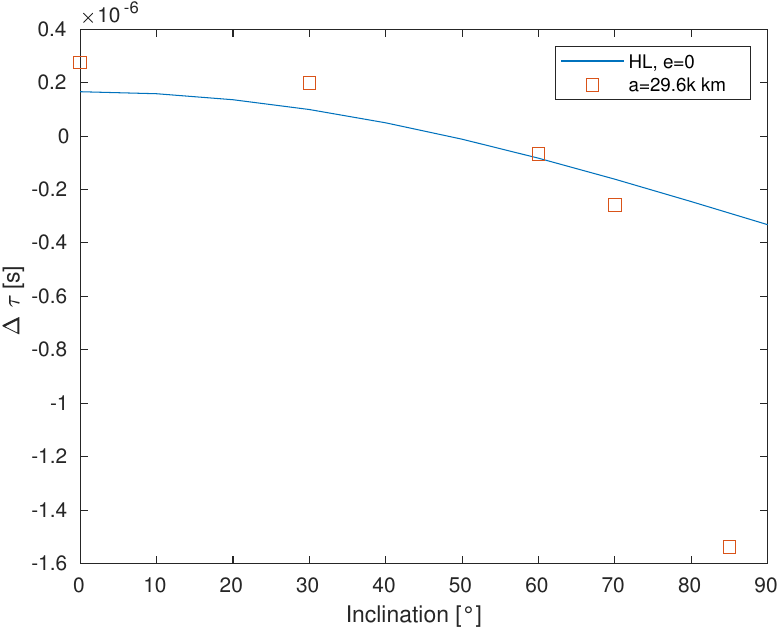}
        \caption{Eccentricity $e = 0$}
        \label{fig:analysis:incliEcc0}
    \end{subfigure}
    \qquad
    \begin{subfigure}{0.4\textwidth}
        \includegraphics[width=\textwidth]{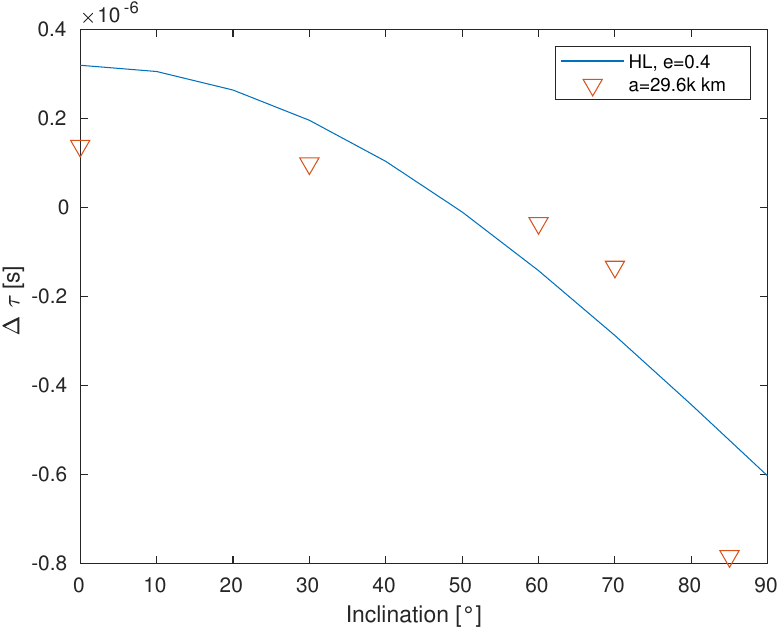}
        \caption{Eccentricity $e = 0.4$}
        \label{fig:analysis:incliEcc04}
    \end{subfigure}     

    \caption{Comparison of theoretical values of the gravitomagnetic clock effect with calculations using the \textit{HPS}. For the simulation, orbits with and without the relativistic corrections for the Lense-Thirring effect were compared. The simulation results are of the same order of magnitude and show the same qualitative behaviour as the theoretical modeled values that cover counter-propagating orbits. The steep drop-off at high inclination angles corresponds to the indefiniteness of the theoretical model at $\varphi = \pi/2$}
    \label{fig:analysis:HL-HPS}
\end{figure}

The gravitomagnetic contribution, due the angular momentum $J$ of the rotating Earth, to a clock's proper time $\tau(\varphi)$ enters our definition of the GMCE in equation~(\ref{eq:gmce-def}). 
For a single clock, it follows that
\begin{align}
    \Delta \tau_{\text{gm}}(\varphi) := \tau(\varphi) - \Hat{\tau}(\varphi) \, ,
    \label{eq:single_sat_diff}
\end{align}
where $\Hat{\tau}(\varphi)$ is the clock's proper in the limit $J=0$.
As observed in Suppl.\ Equation (\ref{supp:eq:dTdphi}), the leading order contribution of $J$ enters through the $dT/d\varphi$ term in Eq.\ \eqref{integral_phi} rather than directly through the metric component $g_{Ti}$ in the integral. 
Since $\Hat{\tau}(\varphi)$ cannot be obtained from the available clock and orbit products, we use the \textit{High Performance Satellite Dynamics Simulator}\cite{listModellingSolarRadiation2015, listHighPrecisionOrbit2021} (\textit{HPS} for short) developed at ZARM and DLR institutes. 
It allows to simulate satellite orbits in a realistic space environment while implementing post-Newtonian corrections (w/o frame dragging). 
To determine whether the \textit{HPS} can reproduce the expected effects, we investigate geodesic orbits, comparable to those in the first theoretical descriptions of the GMCE. 
The following calculations assume a monopolar gravitational field and exclude non-gravitational perturbations. 
To get an approximation for the satellite orbit without the contribution of $ J $, two different orbits are simulated: One with only the Schwarzschild correction taken into account and one also including the Lense-Thirring correction. 
The reasoning behind this is that the angular momentum $ J $ enters the latter for the satellite acceleration directly\cite{petitIERSTechnicalNote}:
\begin{align}
    \Delta \ddot{\vec{r}}_{LT} = (1+\gamma) \frac{GM_E}{c^2 \cdot r^3}\left[ \frac{3}{r^2}(\vec{r} \times \dot{\Vec{r}}) (\Vec{r} \cdot \Vec{J}) + (\dot{\Vec{r}} \times \Vec{J} )\right]  
\end{align}
These orbit models thus provide us with a direct means to get different values for $dT/d\varphi$ depending on the influence of $ J $ and, thus, different elapsed proper times of the respective satellite clock following equation~(\ref{integral_phi}). 
Realistic values of Earth's $J$ and $M$ are used. 
Calculating the gravitomagnetic contribution to clock's proper time, we find for circular equatorial orbits as well as eccentric and inclined geodesics that the effect is reproduced at the correct order of magnitude.
In the latter case the orbital parameters of some sample nominal \textsl{Galileo} satellite are used.
Some more results using different orbital parameters are listed in Table~\ref{tab:analysis:deltaTau_comp} and depicted in Figure~\ref{fig:analysis:HL-HPS}.

As can be seen, the results from the simulated data are in general at the same order of magnitude as the theoretically expected values and roughly follow the theoretical results when taking a look at the derivatives regarding the inclination and eccentricity of the orbit. 
Comparing different approaches with the \textit{HPS} simulation, it shows that taking counter-revolving orbits reproduces the dynamic of the theoretical results in a better way, but the absolute values greatly depend on the radius of the orbit. 
This is not so much the case with the approach of cancelling the acceleration caused by $ J $ via subtracting the two orbits with different levels of relativistic corrections. 
The reasons for the difference in absolute values and the dependency of the orbital radius might be placed in the numerical calculations of the orbit simulation as generally speaking simulations tend to decrease in precision with increased simulation time. 
This reasoning is supported by the declining difference between theory and simulation with declining radial distance, i.e., shorter simulation times. 
The good agreement between theoretical behaviour and simulation results is also displayed in Figure~\ref{fig:analysis:HL-HPS}.

This method could now be used to simulate the real satellite orbits and then compare the calculated proper times with the ones computed from the satellite clock products. 
Another approach would be to use the generalized formulation of the effect\cite{hackmann_generalized_2014} combined with our expanded incremental description of satellite proper time to possibly find satellite constellations in which the gravitoelectric terms are canceled out. Since the leading order of the GMCE is in the $dT/d\varphi$ term, this requires an ultra-precise orbit model. 
This however will in any case need to include non-gravitational forces, within which solar radiation pressure is the most relevant one. Solar radiation pressure (SRP) modelling has been and still is one of the factors limiting the precision of orbit products to the level of cm. Additionally, this modelling is usually not done from a physical a priori model as would be required here. Thus, apart from all other factors that also need to be considered, already to obtain such an a priori SRP model to allow for orbit modelling at $\mu$m precision is a highly demanding task.

\begin{table}
    \centering
    \begin{tabular}{cccrr} \toprule
        $i$ [°] & $e$ & $a$ [km] & $ ^{\text{HL}}\Delta\tau$ [ns]   & $ ^{\text{HPS-LT}}\Delta\tau$ [ns]  \\ \midrule
        0       & 0   & 29,600   & $  166 $ & $   275 $ \\
        30      & 0   & 29,600   & $   99 $ & $   198 $ \\
        85      & 0   & 29,600   & $ -288 $ & $ -1538 $ \\ \midrule
        0       & 0   & 29,600   & $  166 $ & $   275 $ \\
        0       & 0.2 & 29,600   & $  198 $ & $   190 $ \\
        0       & 0.4 & 29,600   & $  319 $ & $   138 $ \\ \midrule
          \bottomrule
    \end{tabular}
    \caption{Comparison of gravitomagetic contribution to proper time, calculated using different approaches. $ ^{HL}\Delta\tau$ is the theoretical model\cite{hackmann_generalized_2014} and $ ^{HPS-LT}\Delta\tau$ the model that aims to cancel the $J$ contribution via the relativistic correction terms.} 
    \label{tab:analysis:deltaTau_comp}
\end{table}

\section*{Mission Scenario, Requirements and Recommendations}

Above we established a general framework for a measurement of the GMCE and evaluated \textsl{Galileo} clock and orbit products as well as orbit simulations using the \textit{HPS} to that end. 
The results show that a measurement of the GMCE from current \textsl{Galileo} clock and orbit products is not yet feasible. 
Thus, in the following we describe what an ideal mission would look like, how it could build on GNSS techniques, and what could be the next steps towards such a mission.

A future mission could make use of technology from GNSS or \textsl{Galileo} in particular, mainly by carrying clocks that build on \textsl{Galileo} heritage and by making use of the extensive worldwide GNSS network to obtain clock and orbit data in the form of precise clock and orbit products. 
At the same time it needs to address some issues that have been raised in the previous section:
\begin{itemize}
    \item[(i)] It has to provide an a priori model of orbital revolution time, that can be compared with the elapsed satellite clock time.
    \item[(ii)] It needs to measure the non-gravitational distortions of the orbit at highest precision. As will be derived below, this needs to be done at least down to accelerations of the order of $ 10^{-12}\,\text{ms}^{-2}$, depending on the orbit choice.
    \item[(iii)] It needs to employ clocks of improved accuracy as compared to existing \textsl{Galileo} PHMs.
\end{itemize}
All of the above requires some significant adaptions with respect to a regular \textsl{Galileo} satellite. 
Thus, it appears likely that a dedicated satellite platform will be required. It will need to carry an upgraded PHM clock, possibly even a clock based on a different technology altogether. 
It will need to be equipped with high performance accelerometers, and maybe even micro-Newton thrusters to realize a drag-free geodesic orbit of the clock. 
At the same time it should be able to operate within the GNSS architecture, i.e., transmit signals to IGS ground stations worldwide.

\subsection*{Requirements on orbits and orbit products}

If a dedicated mission is to incorporate existing GNSS and IGS architecture, the orbit should allow to measure and model the accelerations down to an orbit-dependent level, while having orbital parameters that allow for good ground coverage. 
While a near-polar low Earth orbit offers some advantages regarding the magnitude of the effect and integration times, a medium Earth orbit allows for better incorporation into existing GNSS constellations and will thus be the focus of the following considerations. 

\subsubsection*{Orbit choice and magnitude of the effect}
The choice of inclination and eccentricity of the orbit determines the magnitude of the effect.
This is evident from the generalized expression for a geodesic orbit with inclination $i$ and eccentricity $e$,
 \begin{align}
    \Delta\tau = \tau(2\pi)-\Hat{\tau}(2\pi) &= 4 \pi \frac{J}{M c^2} \frac{(\cos i (3e^2+2e+3)-2e-2)}{(1-e^2)^{\frac{3}{2}}} \, .
    \label{eq:geneffect}
\end{align}
Accordingly, the effect for a circular, equatorial orbit, which is at $\Delta \tau = 69$\,ns, can be significantly enhanced by choosing a large eccentricity $e$ and an appropriate inclination $i$. 
In fact the optimum inclination is for a close to polar orbit at $i \approx 90^{\circ}$ which leads to a factor of two increase over the signal with equatorial orbits, while the effect is suppressed by a factor $0.32$ for a \textsl{Galileo} orbit at $i = 56^{\circ}$ and vanishes for a circular orbit at $i = 48.2^{\circ}$. 
Increasing the eccentricity on the other hand provides an even stronger leverage. 
An eccentricity of $e = 0.6$ already results in an effect enhanced by more than a factor of 4, and $e = 0.9$ would boost the signal by a factor of approximately 40. Clearly though, the eccentricity cannot be chosen arbitrarily large. For simplicity we consider the case of an equatorial orbit with apogee at \textsl{Galileo} altitude ($r_{max} = 29600 $\,km) and a perigee at 1000\,km altitude (i.e., $r_{min} \approx 7370$\,km). This corresponds to an eccentricity of $e = 0.6$ and a signal of $\Delta \tau = 280 $\,ns. By choosing a polar orbit of the same eccentricity this could even be enhanced further up to $\Delta \tau = 430 $\,ns, which is a gain by a factor of approximately 6 over the circular equatorial case. 

On the other hand, we need to take into account that the smaller single orbit effect for a low-Earth orbit (LEO) ($r_{min} \approx 7370$\,km, $T = 6297$\,s) will accumulate by almost a factor of 9 over the time of a single revolution in the eccentric medium-Earth orbit (MEO) ($T = 54300$\,s), which effectively leads to a larger integrated signal over the same time span. All the while the residual offset of the clock frequency will accumulate linearly in time as well, so its effect will be the same per integrated measurement span in both orbits. 

Considering only this aspect of the mission, a circular LEO would be preferred over both an eccentric or a circular MEO. 
However, residual atmospheric drag as well gravitational accelerations from Earth mass multipoles increase significantly towards lower altitudes. Consequently, the decision on the orbit needs to be based on a trade-off that takes into account the magnitude of orbit distortions, the required precision on orbital parameters and the magnitude of the expected signal and its accumulation as just discussed.

First we assess the impact of deviations of the orbit radius $\Delta r$ and the azimuthal deviation $\Delta \varphi$ from a comparison at $2\pi$. For an ideal circular equatorial orbit such deviations will change the result to
\begin{equation}
\tau_+ -\tau_- \approx 4\pi \frac{J}{Mc^2} + \sqrt{\frac{r^3}{G M}}\Delta \varphi + 3\pi \sqrt{\frac{r}{G M}} \Delta r.
\end{equation}
This expression can be generalized to an eccentric, inclined Kepler orbit as well, including also deviations in inclination and eccentricity. For the purpose of estimating orders of magnitude it is sufficient to stick to the special cases of equatorial or (close to) polar orbits though. The resulting requirements for a circular, polar orbit are shown in Table~\ref{tab:orbreq}. 
We also give a simplified estimate translating these requirements on orbit position into accelerations over the time of one orbital revolution. Thereby we assume constant acceleration in either radial or azimuthal direction over the full course of one orbit as a worst case scenario.

\begin{table}[t]
  \centering
  \begin{tabular}{cccccccc} \toprule
  	 orbit & $r$\,[km] & $T [s]$ & $\Delta r$\,[$\mu$m] & $a_r \, [m/s^2$] & $\Delta\varphi$\,[prad]   & $r\times\Delta \varphi$\,[$\mu$m] & $a_{\varphi} \, [m/s^2$] \\ \midrule
  	 LEO   & 7370 & 6297 & 10.68 & $5.38 \times 10^{-13}$ & 13.66  &  100.74 & $5.08 \times 10^{-12}$ \\ 
  	 MEO & 29600 & 50681 & 5.34 & $4.2 \times 10^{-15}$ & 1.7 &  50.26 & $4.0 \times 10^{-14}$ \\
  	   \bottomrule
  \end{tabular}
  \caption{Requirements derived from the top level mission goal on a circular equatorial orbit. A measurement at 10\% precision corresponds to a maximum allowed uncertainty in $\Delta \tau$ of 6.9\,ns. Note that the above accelerations are derived under the assumption that they act continuously along r or in azimuthal direction over the full orbit. Thus, these estimates do not reflect expected effects of averaging, which should relax requirements with respect to several sources.}
 \label{tab:orbreq}	
\end{table}

\subsubsection*{Gravitational orbit effects}
Our estimates in Table~\ref{tab:orbreq} suggest that accelerations down to the $10^{-15}$\,ms$^{-2}$ level need to be taken into account in case of a MEO orbit and $10^{-13}$\,ms$^{-2}$ for a LEO. 
For Earth's gravitational potential this means that mass multipole contributions down the level of $J_{11}$ need to be taken into account in a MEO and even more for a LEO, see Suppl.~Table~\ref{supp:tab:zonal_terms_acc}. The accuracy at which these contributions are actually known from the GOCE/GRACE data \cite{EGM2008} is also given there. 
For $J_2$ and a MEO this precision is at $\sigma \approx 5\times 10^{-13}$\,ms$^{-2}$, which is expected to be the dominating uncertainty in modelling these gravitational distortions, and which would still be about two orders of magnitude too large. 
However, the estimates we have given in Table~\ref{tab:orbreq} are only worst case estimates and a more detailed modelling may show that this requirement can be relaxed, e.g., due to averaging or partial compensation of these effects. 

Contributions from other solar system bodies are at a level of $10^{-6}$\,ms$^{-2}$ for Sun and Moon, and at $10^{-10}$\,ms$^{-2}$ for the planets. 
These contributions also need to be included in the modelling down to Pluto, relying ephemerides as well as relevant spherical harmonic coefficients as well.

\subsubsection*{Non-gravitational orbit distortions}

The dominating non-gravitational acceleration on a MEO is due to SRP at a level of $10^{-7}$\,ms$^{-2}$. 
This is followed by thermal radiation pressure from the satellite approximately an order of magnitude below. 
Earth albedo finally contributes on the order of~$10^{-9}$\,ms$^{-2}$, while atmospheric drag at this altitude is negligible. 
For the considered LEO, atmospheric drag is still relevant possibly up to the  SRP level and also the effect from Earth albedo is increased by about one order of magnitude. 
All of these effects of course depend strongly on the varying attitude of the satellite with respect to Sun and Earth, on its specific material properties, and ageing thereof, as well as internal dissipation of the payload. 
Modelling these effects down to the required levels, i.e., below $10^{-12}$\,ms$^{-2}$ is certainly not feasible, even with an anticipated significant progress in modelling techniques.

Contrary to gravitational accelerations however, non-gravitational accelerations onto the satellite can be measured by an on-board accelerometer. 
These measurements can be used to either take them into account in the modelling or to compensate their effect in the first place, realizing a near geodesic orbit by drag-free control. 
With the MICROSCOPE mission \cite{MICROSCOPE} a drag-free satellite has been flown on a LEO with residual accelerations down to $1.5 \times 10^{-12}$\,ms$^{-2}$Hz$^{-1/2}$. 
While challenging, it thus still appears feasible at least in principle to achieve the necessary control of the effects to the required level.

\subsection*{Requirements on clocks and clock products}
To deduce a requirement on the clock, we need to specify the magnitude of the effect and, thus, the orbit under consideration. 
Picking the optimum case of a circular polar orbit results in an effect of $\Delta \tau = 137$\,ns, which we want to observe at the 10\% level. 
This translates into a requirement on the drift of the satellite clock during an orbital period $T$ of $\frac{\Delta t}{t}< 13.7\,\text{ns} / T$.  
Note that for an equatorial orbit it reduces by a factor of 2, and for arbitrary inclination may decrease further.

\begin{table}
  \centering
  \begin{tabular}{ccccc} \toprule
  	 orbit & $r$\,[km] & $T$\,[s] & $\Delta \nu/\nu$    \\ \midrule
  	 LEO   & 7370 & 6297 & $< 2.2 \times 10^{-12}  $    \\ 
  	 MEO & 29600 & 50681 & $< 2.7 \times 10^{-13} $    \\
  	   \bottomrule
  \end{tabular}
  \caption{Requirements on the clock for polar orbits of different radii, all exhibiting an effect of $\Delta \tau = 137$\,ns to be measured at the 10\% level. Stated is the required clock accuracy and repeatability of its frequency. This implies that also the relative clock stability at time scale of orbital revolution needs to be reduced below this level. }
 \label{tab:clockreq}	
\end{table}

Depending on the choice of orbit radius and the resulting orbital revolution period $T$ this translates into requirements on the knowledge of the satellite clock's frequency $\nu_s$ relative to the reference clock frequency $\nu_r$ as listed in Table~\ref{tab:clockreq} with $ \Delta\nu/\nu = \frac{\nu_s-\nu_r}{\nu_r}$. 
Of course one might argue that these respective clock biases are all estimated when processing the clock and orbit data, thus the relative offsets are all known already. 
But this estimation process will then simply estimate technical drifts and offsets together with the effect we want to measure. 
For discriminating the gravitomagnetic contribution within the estimated biases, an a priori knowledge of two frequencies is required.

The analysis of clock and orbit products so far shows drifts at the order of 1\,$\mu$s per 50000\,s. 
A part of this drift is most likely due to a deliberate frequency offset.
For a dedicated mission, we expect that it should be possible to model and reduce this drift to the limit set by the uncertainty of the absolute frequency of current \textsl{Galileo} PHMs, at approximately 100\,ns per orbit. 
For a measurement on a MEO orbit, this is approximately one order of magnitude above the requirement. 
Consequently, a future measurement should use clocks that are intrinsically more accurate, or use specific operation and monitoring of the clocks to overcome this technical limit.

\section*{Conclusion}

In this work we provided an incremental definition of the gravitomagnetic clock effect that allows to model the effect along arbitrary worldlines. 
We find, as reported by other authors before, that the gravitomagnetic contribution to the proper time $\tau$ appears only at the order $1/c^4$ as $\sim GJ/c^4r$. 
A measurement of this term would require a clock at a frequency accuracy of $\Delta\nu/\nu \sim 10^{-22}$, which is more than three orders of magnitude below what the most accurate clocks to date can achieve. 
The leading order effect described in this work enters via the angular rate of change $dT/d\varphi$ into the proper time integral. 
In order to discriminate the gravitomagnetic contribution to this term bases on existing GNSS data, the orbit data would need to be corrected by accurate models of all orbit distortions down to $\mu$m precision. 
Additionally, most GNSS satellites operate on unfavorably inclined orbits for which the effect is suppressed by a factor $\sim 0.3$. 
A dedicated mission would thus benefit from different inclinations, mostly lower inclinations to improve integration with existing systems.
Modelling the gravitational distortions at the required precision in a dedicated mission is at the edge of what is possible today based on data from the GRACE and GOCE missions. 
Increasing the orbit radius to reduce modelling requirements on the higher multipole moments is recommended, although this will slow down the accumulation of the signal. 
Comparing to a test of general relativistic frame dragging using the LAGEOS satellites suggests that the modelling for the proposed mission is at least one order of magnitude more demanding.
One advantage would be that a dedicated mission could acquire data over an integration time of several years. 
This allows to relax the required precision of the orbit model to the cm level, which is currently achievable. 
How the averaging effects introduced by the long integration time influence the measurement is yet to be modelled in more detail.

\noindent Based on the above, we can state that a measurement of the gravitomagnetic clock effect is highly demanding, but might just be within reach of what is possible based on current technology. 
An important theoretical aspect yet to be investigated more thoroughly is to what extent this measurement tests similar or different aspects of gravitomagnetism as compared to other missions such as LAGEOS.
 
Should such a dedicated mission be implemented, it will definitely need to build on the unique ability of the GNSS network to provide accurate data on orbits and clocks involved. 
Thus, relevant demonstrations and in orbit verification of the necessary steps towards this measurement might already be done wherever the opportunity arises. 

\bibliography{gmce_lit}

\section*{Acknowledgements}

Funded by the Deutsche Forschungsgemeinschaft (DFG, German Research Foundation) – Project-ID 434617780 – SFB 1464.\\
Funded by the Deutsche Forschungsgemeinschaft (DFG, German Research Foundation) under Germany’s Excellence Strategy – EXC-2123 QuantumFrontiers – 390837967, and through the Research Training Group 1620 ``Models of Gravity''.\\
This work was also supported by the German Space Agency DLR with funds provided by the Federal Ministry of Economics and Technology (BMWi) under Grant No.\ DLR 50WM1547 and the DLR institute for Satellite Geodesy and Inertial Sensing.

We gratefully acknowledge support by the European Space Agency ESA.
This work is supported by the German Space Agency (DLR) with funds provided by the Federal Ministry of Economic Affairs and Climate Action (BMWK) under FKZ: 50WM1960

\section*{Author contributions statement}

S. Herrmann, C. Lämmerzahl, J. Ventura-Traveset and L. Mendes proposed the study on which this work is based. S. Herrmann, C. Lämmerzahl, E. Hackmann, D. Philipp and J. Scheumann realized the theoretical framework and the data analysis described here. B. Rievers contributed methods for orbit modelling.

\section*{Additional information}

The authors declare no competing interests.
The corresponding authors can grant access to code and data on reasonable request.

\label{mylastpage}
\clearpage
\section*{Supplemental material}

\renewcommand{\figurename}{Supplementary Figure}
\renewcommand{\tablename}{Supplementary Table}

\renewcommand{\theequation}{S\arabic{equation}}
\renewcommand{\thepage}{S\arabic{page}}

\setcounter{page}{1}
\setcounter{equation}{0}
\setcounter{figure}{0}
\setcounter{table}{0}

\rfoot{\small\sffamily\bfseries\thepage/\pageref{LastPage}}

\subsection*{Equation of motion}
The equation of motion in GR is the geodesic equation,
\begin{align}
    \dfrac{d^2 x^\mu}{d\tau^2} + \Gamma^{\mu}{}_{\nu\sigma} \dfrac{dx^\nu}{d\tau}\dfrac{dx^\sigma}{d\tau} = 0 \, .
    \label{supp:eq:geodesic}
\end{align}
Using a computer algebra program, we have calculated all non-vanishing Christoffel symbols $\Gamma^{\mu}{}_{\nu\sigma}$ for the metric under consideration in the post-Newtonian expansion.

Proper time $\tau$ is genuinely defined by the clock postulate and related to the normalization of an observer's four-velocity, which in our case reads
\begin{align}
    -c^2 = g(u,u) = g_{TT} (u^T)^2 + 2g_{T\varphi} u^T u^\varphi + g_{ij} u^i u^j \, ,
\end{align}
where $u^\mu = dx^\mu / d\tau$.
Thus, only for proper time-parametrized motion, the normalization takes this form.
A clock that shows proper time is called a standard clock and all measurements are in agreement with the fact that atomic clocks are standard clocks.
Standard clocks can indeed be characterized operationally \cite{perlickCharacterizationStandardClocks1987}.
The proper time increment that elapses during the motion along a clock's worldline can be calculated from
\begin{align}
    -c^2 d\tau^2 = g_{TT} dT^2 + 2 g_{T\varphi} dT d\varphi + g_{ij} dx^i dx^j \, .
\end{align}
Note that in our convention $g_{TT}<0$ and $g_{T\varphi} < 0$.
Integrating this expression between two events on the worldline yields the elapsed proper time.
For the integration, we need to parametrize the worldline.
Let $q$ be some parameter which we use in this respect. 
Then
\begin{align}
   \dfrac{d\tau}{dq} = \pm \dfrac{1}{c} \sqrt{-g_{TT} \left( \dfrac{dT}{dq} \right)^2 - 2 g_{T\varphi} \dfrac{dT}{dq}\dfrac{d\varphi}{dq} - g_{ij} \dfrac{dx^i}{dq} \dfrac{dx^j}{dq} } \, ,
\end{align}
and we may choose the upper sign to have the proper time increase in the direction of increasing $q$.
The integration formally yields
\begin{align}
    \tau - \tau_0 = \int_{q_0}^q \dfrac{1}{c} \sqrt{-g_{TT} \left( \dfrac{dT}{dq} \right)^2 - 2 g_{T\varphi} \dfrac{dT}{dq}\dfrac{d\varphi}{dq} - g_{ij} \dfrac{dx^i}{dq} \dfrac{dx^j}{dq} } \, dq \, .
\end{align}
One obvious choice is to paramterize the motion by coordinate time. Then
\begin{align}
    \tau - \tau_0 = \int_{T_0}^T \dfrac{1}{c} \sqrt{-g_{TT} - 2 g_{T\varphi} \dfrac{d\varphi}{dT} - g_{ij} \dfrac{dx^i}{dT} \dfrac{dx^j}{dT} } \, dT \, ,
\end{align}
which can be simplified by formally introducing the coordinate velocity
\begin{align}
    v^\mu := \dfrac{dx^\mu}{dT} \, , \quad (v^\mu) = (1,dx^i/dT) = (1,v^i) = (1,\vec{v}) \, .
\end{align}
Thus, we have
\begin{align}
    \tau - \tau_0 = \int_{T_0}^T \dfrac{v}{c} \, dT \, , \quad v = \sqrt{-g_{\mu\nu} v^\mu v^\nu} \, .
\end{align}
For geodesics, the velocity $v$ is a function of position, mass multipole moments, and angular momentum, i.e, ${v = v(x,M_{ij},J)}$.
We may choose the initial conditions such that $\tau_0 = 0$ at $T_0 = 0$.
Note that for geodesics in the stationary spacetime, $v/c$ is related to the total (dimensionless) energy of the system, which is conserved such that for geodesic orbits
\begin{align}
    \tau = \dfrac{1}{C_E} \int_0^T \left( g_{TT} - g_{T\varphi} \dfrac{d\varphi}{dT} \right) dT \, ,
\end{align}
where $C_E$ is the conserved energy constant of motion.
On the geoid we have $v = c$ such that, by construction, proper time and coordinate time coincide.

Another possible choice is to use the azimuthal angle $\varphi$ as a curve parameter. Thus,
\begin{align}
    c \dfrac{d\tau}{d\varphi} = \pm \sqrt{-g_{TT} \left( \dfrac{dT}{d\varphi} \right)^2 - 2 g_{T\varphi} \dfrac{dT}{d\varphi} - g_{ij} \dfrac{dx^i}{d\varphi} \dfrac{dx^j}{d\varphi} } \, .
\end{align}
Integration formally results in (assuming $\tau = 0$ at $\varphi = 0$)
\begin{align}
    c \tau(\varphi) = \pm \int_0^{\varphi} \sqrt{-g_{TT} \left( \dfrac{dT}{d\varphi} \right)^2 - 2 g_{T\varphi} \dfrac{dT}{d\varphi} - g_{ij} \dfrac{dx^i}{d\varphi} \dfrac{dx^j}{d\varphi} } \, d\varphi \, . \label{supp:eq:integral}
\end{align}
The sign needs to be chosen according to the direction of revolution.
Up to here, the results are general and no special type of orbit is assumed. 

\subsection*{Details on the parameter $ \alpha $}
The post-Newtonian expansion of the factor of proportionality $ \alpha $\cite{hackmann_generalized_2014}, which depends on the orbital elements, is
\begin{equation}
\alpha \approx - \frac{d_1^{\frac{3}{2}}}{d_2^{\frac{3}{2}}} 
               - \frac{3 d_1^{\frac{1}{2}}}{2 d_2^{\frac{5}{2}}} \left[ 
                 \frac{d_1(1+e_2^2)}{1-e_2^2} - \frac{d_2(1+e_1^2)}{1-e_1^2}
                \right] \frac{G M}{c^2},
\label{supp:eq:alpha_PN_approx}
\end{equation}
and, accordingly, the generalized GMCE is
\begin{equation}
\Delta \tau_{\text{gm}} \approx \frac{J}{Mc^2} \left[
                                    s_1 \frac{2\pi(\cos i_1 (3e_1^2+2e_1+3)-2e_1-2)}{(1-e_1^2)^{\frac{3}{2}}} 
                                  - s_2 \frac{d_1^{\frac{3}{2}}}{d_2^{\frac{3}{2}}} \frac{2\pi(\cos i_2 (3e_2^2+2e_2+3)-2e_2-2)}{(1-e_2^2)^{\frac{3}{2}}}
                                  \right],
\label{supp:eq:d_tau_first_order}                                  
\end{equation}
where, $e_{\text{i}}, i_{\text{i}}, d_{\text{i}}$ are defined for the orbits $i = 1,2$ as above for equation (\ref{eq:generalized_tau}) and $s_{\text{j}}$ equals to $\pm 1$ depending on the revolution sense.

\subsection*{Pinpointing frame dragging contribution of $ J $}
We consider again the integral
\begin{align}
    \tau(\varphi) = \pm \dfrac{1}{c} \int_0^{\varphi} \sqrt{-g_{TT} \left( \dfrac{dT}{d\varphi} \right)^2 - 2 g_{T\varphi} \dfrac{dT}{d\varphi} - g_{ij} \dfrac{dx^i}{d\varphi} \dfrac{dx^j}{d\varphi} } \, d\varphi \, . 
\end{align}
that describes elapsed proper time on a general orbit, which is paramatrized by the azimuthal angle.
The task is to pinpoint where leading-order contributions of gravitomagnetic effects come into play.
We know that the metric functions are given to the order
\begin{subequations}
\begin{align}
    g_{TT} &\sim \mathcal{O}\big( c^{2} + U + U / c^{2} \big) \, , \\
    g_{T\varphi} &\sim \mathcal{O}\big( G J /c^{2} \big) \, , \\
    g_{rr} &\sim \mathcal{O}\big(1 + U / c^{2} \big) \, , \\
    g_{\vartheta\vartheta} &\sim \mathcal{O}\big( 1 + U / c^{2} \big) \, , \\
    g_{\varphi\varphi} &\sim \mathcal{O}\big( 1 + U / c^{2} \big) \, .
\end{align}
\end{subequations}
We further have 
\begin{align}
\dfrac{dT}{d\varphi} = \left. \dfrac{dT}{d\varphi} \right|_\text{N} + c^{-2} \left. \dfrac{dT}{d\varphi} \right|_\text{pN} + \mathcal{O} \big( c^{-3} \big) \, ,
\end{align}
where $\left. \dfrac{dT}{d\varphi} \right|_\text{pN}$ is a function of position and of $(U,\beta,\gamma,J)$.
We formally write
\begin{align}
    \tau(\varphi) = \pm \int_0^{\varphi} \sqrt{\kappa_0 + \dfrac{\kappa_2}{c^2} + \dfrac{\kappa_4}{c^4} } \, d\varphi \, ,
\end{align}
where $\kappa_i$ depend on $\varphi, U$ and $\kappa_2,\kappa_4$ depend on $\beta,\gamma,J$ as well.
Series expanding the integrand leads to
\begin{align}
     \tau(\varphi) = \pm \int_0^{\varphi} \left( \sqrt{\kappa_0} + \dfrac{1}{c^2} \left( \dfrac{\kappa_2}{2 \sqrt{\kappa_0}} \right)  \right) d\varphi \, .
\end{align}
Note that $\kappa_0$ is the Newtonian term (Kepler's period if the situation was spherically symmetric), and $\kappa_2$ is the post-Newtonian contribution to the result.

We find
\begin{subequations}
\begin{align}
\kappa_0 &= \left( \left. \dfrac{dT}{d\varphi} \right|_\text{N} \right)^2 = 1/\omega_0^2 = \left( \dfrac{GM}{r^3} \left( 1+\bar{U} - r^2\bar{U}_{,r} \right)\right)^{-1} \, , \\
    \kappa_2 &= \sqrt{\kappa_0} \left. \dfrac{dT}{d\varphi} \right|_\text{pN} + 2(U - U^*_0) \kappa_0 - \left. g_{ij} \dfrac{dx^i}{d\varphi} \dfrac{dx^j}{d\varphi} \right|_\text{N} \, .
\end{align}  \label{supp:eq:dTdphi}
\end{subequations}
Thus, we see that at leading order $J$ enters the expression for proper time via the term $\left. \dfrac{dT}{d\varphi} \right|_\text{pN}$ in $\kappa_2$ and not due to $g_{T\varphi}$ in the integral that we started with.
Now one needs to make a model of $\kappa_2(J=0) := \hat{\kappa}_2$ and, thereupon, a model of $\hat{\tau}$ that can be subtracted from observational data.

\subsection*{Gravitational and non-gravitational orbit distortions}
Table~S\ref{supp:tab:zonal_terms_acc} displays the accelerations caused by the zonal terms of the multipole moments relevant to GALILEO orbits.
\begin{table}[h]
    \centering
    \begin{tabular}{crc|rc} \toprule
        Term & $ \Vec{a}_{\text{MEO}} $ [ms$^{-2}$] & $ \sigma_{\text{EGM 2008}} $ [ms$^{-2} $] & $ \Vec{a}_{\text{LEO}} $ [ms$^{-2}$] & $ \sigma_{\text{EGM 2008}} $ [ms$^{-2} $] \\ \midrule
        $J_2$  &  $ -3.07 \times 10^{-05} $ & $ 4.74 \times 10^{-13} $ & $ -7.98 \times 10^{-3} $ & $ 1.23 \times 10^{-10} $ \\
        $J_3$  &  $  1.74 \times 10^{-08} $ & $ 1.04 \times 10^{-13} $ & $  1.82 \times 10^{-5} $ & $ 1.09 \times 10^{-10} $ \\
        $J_4$  &  $  2.65 \times 10^{-09} $ & $ 2.17 \times 10^{-14} $ & $  1.11 \times 10^{-5} $ & $ 9.12 \times 10^{-11} $ \\
        $J_5$  &  $  8.71 \times 10^{-11} $ & $ 3.70 \times 10^{-15} $ & $  1.47 \times 10^{-6} $ & $ 6.22 \times 10^{-11} $ \\
        $J_6$  &  $ -4.78 \times 10^{-11} $ & $ 6.49 \times 10^{-16} $ & $ -3.24 \times 10^{-6} $ & $ 4.39 \times 10^{-11} $ \\
        $J_7$  &  $  7.11 \times 10^{-12} $ & $ 1.21 \times 10^{-16} $ & $  1.93 \times 10^{-6} $ & $ 3.29 \times 10^{-11} $ \\
        $J_8$  &  $  9.41 \times 10^{-12} $ & $ 2.35 \times 10^{-17} $ & $  1.03 \times 10^{-6} $ & $ 2.57 \times 10^{-11} $ \\
        $J_9$  &  $  1.28 \times 10^{-13} $ & $ 4.67 \times 10^{-18} $ & $  5.60 \times 10^{-7} $ & $ 2.05 \times 10^{-11} $ \\
        $J_{10}$ &  $  5.76 \times 10^{-14} $ & $ 9.52 \times 10^{-19} $ & $  1.01 \times 10^{-6} $ & $ 1.68 \times 10^{-11} $ \\
        $J_{11}$ &  $ -1.29 \times 10^{-14} $ & $ 1.93 \times 10^{-19} $ & $ -9.12 \times 10^{-7} $ & $ 1.36 \times 10^{-11} $ \\
        $J_{12}$ &  $  2.16 \times 10^{-15} $ & $ 4.05 \times 10^{-20} $ & $  6.13 \times 10^{-7} $ & $ 1.15 \times 10^{-11} $ \\
        $J_{13}$ &  $  5.74 \times 10^{-16} $ & $ 8.63 \times 10^{-21} $ & $ -3.30 \times 10^{-7} $ & $ 9.85 \times 10^{-12} $ \\
    \bottomrule
    \end{tabular}
    \caption{Accelerations for a Galileo orbit (MEO @ r = 29600\,km, LEO @ r = 7370\,km) caused by zonal terms of multipole moments. The standard deviation $\sigma$ was calculated from the standard deviation of the coefficients as stated by the \textit{EGM 2008} \cite{EGM2008}. (Not all significant digits are shown here.)}
    \label{supp:tab:zonal_terms_acc}
\end{table}

\end{document}